\documentclass{iopconfser}
\usepackage{graphicx} 
\usepackage{orcidlink}
\usepackage{amsmath,amssymb}
\begin{document}

\title{Aspects of Geometrodynamics in the Jordan and Einstein Frames.}

\author{Gabriele Gionti, S.J. $^{1}$ \orcidlink{0000-0002-0424-0648}}

\affil{ $^{1}$ Specola Vaticana (Vatican Observatory), V-00120, Vatican City State

Vatican Observatory Research Group, Steward Observatory, The University of Arizona, 933 North Cherry Avenue, Tucson, Arizona 85721, USA and 

INFN, Laboratori Nazionali di Frascati, via E. Fermi 40, 00044 Frascati, Italy}

\email{ggionti@specola.va}

\begin{abstract}
We will summarize recent results on the Hamiltonian equivalence between the Jordan and Einstein frames based on the analysis of Brans-Dicke theory for both cases $\omega\neq -\frac{3}{2}$  and $\omega =-\frac{3}{2}$.

We will introduce and perform ADM analysis for spherically symmetric solutions of gravity. We will discuss with particular care the problem of the boundary terms to be introduced in the general case of spherical symmetry. These two frames are connected through a Hamiltonian canonical transformation on the reduced phase space obtained by gauge fixing the lapse and the radial shift functions. We introduce and discuss two static solutions (Fisher, Janis, Newman and Winicour solution in the Einstein frame and Bocharova-Bronnikov-Melnikov-Bekenstein black hole solution in the Jordan frame) 
\end{abstract}

\section{\label{Jordan-Einstein}Introduction}

It is fairly well known that we never measure in physics absolute quantities, but ratios of absolute quantities. In fact, we need to define a unit of measurement $u$ and determine how many times this unit of measurement is contained in the quantity we want to measure. For example, suppose we work in natural units where the mass has the dimension of the inverse of length \cite{Dicke}.  Be $m_p$ the mass of the proton with respect to the unit of measurement ${m}_u$ and rescale the unit of measurement by a factor ${\lambda}^{-1}$ (we stress that both in \cite{Dicke} and \cite{Faraoni2006} $\lambda(x)$ is a local function), that is, ${\tilde m}_u={\lambda}^{-1}{m}_u$, this implies that in this new unit of measurement ${\tilde m}_p={\lambda}^{-1}{m}_p$ and the ratio \cite{Faraoni2006} stays constant. 

\begin{equation}
\frac{{\tilde m}_p}{{\tilde m}_u}=\frac{{\lambda}^{-1}{m}_p}{{\lambda}^{-1}{m}_u}=\frac{m_p}{m_u}\,.
\label{physicalequiv}
\end{equation}

This rescaling appears more intuitive repeating these reasoning on length scales. In fact, in natural units,  \cite{Dicke} the above rescaling on the masses implies a length rescaling $d\tilde{s}=\lambda ds$. Now we have that $ds=\left(g_{ij}dx^{i}dx^{j}\right)^{\frac{1}{2}}$. Holding the same coordinates, the metric coefficients transform  

\begin{equation}
{\tilde g}_{\mu\nu}={\lambda}^{2} g_{\mu\nu}\,\,.
\label{conformal}
\end{equation}

Therefore \cite{Dicke} invariance of the physical observables under rescaling of units of measurements implies invariance under Weyl(conformal) rescaling of the coefficients of the metric tensor. The starting frame is called {\it Jordan} frame and the conformally transformed {\it Einstein} frame. One observable can be measured either in the Jordan or the Einstein frame. The two measures are related through a power of the conformal factor according to the dimension of the observable in natural units (in the example of the mass of the proton, the measure of the observable in the Einstein frame is equal to the measure of the observable in the Jordan frame multiplied by the inverse of the square root of the conformal factor). 

\section{Scalar-tensor theory}
Nowadays, the general procedure \cite{Faraoni2006} is to start with a scalar-tensor theory action \cite{Dyer}  with the Gibbons-Hawking-York (GHY) boundary term \cite{gibbons&hawking} \cite{york1} \cite{york2} in what is called {\it Jordan frame}

\begin{eqnarray}
S&=&\int_{M}d^{n}x{\sqrt{-g}}\left(f(\phi)R-\frac{1}{2}\lambda(\phi)g^{\mu\nu}\partial_{\mu}\phi\partial_{\nu}\phi -U(\phi)\right) \nonumber \\
&+&2\int_{\partial M}d^{n-1}{\sqrt{h}}f(\phi)K \,.
\label{scalartensor}
\end{eqnarray}

where $f(\phi)$ is a generic function of $\phi$ as well as $\lambda(\phi)$, $K$ is the trace of the extrinsic curvature. This theory represents a generic scalar field non-minimally coupled to the gravitational field. If we perform the variation with respect the metric $g_{\mu\nu}(x)$ and set its variation $\delta g_{\mu\nu}(x)=0$ on the boundary,  we get the equations of motion for it 

\begin{equation}
f( \phi)\left(R_{\mu\nu}-\frac{1}{2}g_{\mu\nu}R\right)+g_{\mu\nu} \Box f(\phi)-\nabla_{\mu}\nabla_{\nu}f(\phi)=T^{\phi}_{\mu\nu}, 
\label{Einsteinequiv}
\end{equation}

where 
\begin{equation}
T^{\phi}_{\mu\nu}=\lambda(\phi)\partial_{\mu} \phi \partial _{\nu} \phi -\frac{1}{2}g_{\mu\nu}\lambda(\phi)g^{\alpha\beta}\partial_{\alpha} \phi \partial _{\beta} \phi -U(\phi)\,.
\label{tensorimpu}
\end{equation}

Variations respect to $\phi(x)$ and imposing these variations are zero at the boundary $\delta \phi(x)=0$ provide equation of motion for the scalar field $\phi (x)$

\begin{equation}
f'(\phi)R+\frac{1}{2}\lambda'(\phi)(\partial \phi)^{2}+\lambda(\phi) \Box \phi -U'(\phi)=0\;\;\;\,.
\label{eqaphi}
\end{equation}

In the literature one passes from the Jordan to the { \it Einstein frame} \cite{Dicke} \cite{Faraoni2006} through a Weyl transformation of the metric, above mentioned, which now, for convenience, we choose to be

\begin{equation}
{\tilde g}_{\mu\nu}=\Big(16\pi G f(\phi)\Big)^{\frac{2}{n-2}}g_{\mu\nu}\;, 
\label{Weyltrans}
\end{equation}

${\tilde g}_{\mu\nu}$ being the metric tensor in the Einstein frame. In the Einstein frame the action \eqref{scalartensor} becomes 

\begin{eqnarray}
S&=&\int_{M}d^{n}x{\sqrt{-{\tilde g}}}\left(\frac{1}{16\pi G}{\tilde R}-A(\phi){\tilde g}^{\mu\nu}\partial_{\mu}\phi\partial_{\nu}\phi -V(\phi)\right) \nonumber \\
&+&\frac{1}{8\pi G}\int_{\partial M}d^{n-1}{\sqrt{\tilde h}}{\tilde K}
\label{scalartensorEF},
\end{eqnarray}

where 

\begin{eqnarray}
A(\phi)&=&\frac{1}{16\pi G}\left(\frac{\lambda(\phi)}{2f(\phi)}+\frac{n-1}{n-2}\frac{(f'(\phi))^2}{f^2(\phi)}\right), \nonumber \\ 
V(\phi)&=&\frac{U(\phi)}{[16\pi G f(\phi)]^{\frac{n}{n-2}}}.
\label{AandV}
\end{eqnarray}

\noindent varying this equation respect to ${\tilde g}^{\mu\nu}$ we get Einstein Equations minimally coupled to the scalar field $\phi$ and varying respect to $\phi$ we get the equation for $\phi(x)$. It is always assumed that, e.g.\cite{Dicke} \cite{Faraoni2006}, if the couple $(g_{\mu\nu}(x),\phi(x))$ is a solution of the equations of motion in the Jordan frame, then, by construction, $(\tilde{g}_{\mu\nu}(x,\phi), \phi(x))$ is solution of the equations of motion in the Einstein frame. Therefore, in the litterature, we find the remark that the two frames are physically equivalent provided the scaling relations among observables quantities in the two frames \cite{Dicke} \cite{Faraoni2006} \cite{Cho1992}. 
\section{Hamiltonian Canonical Transformation}
The previous statement, in our opinion, is far to be crystal clear. One way to check it is to pass to the Hamiltonian formalism and see whether the transformation from the Jordan to the Einstein frames is a Hamiltonian canonical transformation \cite{GiontiSJ:2023tgx}. That is to say, if the couple $(g_{\mu\nu}(x),\phi(x))$ is a solution of the equations of motion in the Jordan frame, then, by construction,  $(\tilde{g}_{\mu\nu}(x,\phi) \phi(x))$ is solution of the equations of motion in the Einstein frame. Once we proved this, a Hamiltonian canonical transformation exhibits a mathematical equivalence,, not a physical equivalence.  

We have found, in the literature, that sometimes the Hamiltonian transformation from the Jordan to the Einstein frames is assumed Hamiltonian canonical without proving this statement \cite{Garay1992}. They start from the following relations
\begin{equation}
{\widetilde{q}}_{ik}=\Phi q_{ik},\,\,
\phi = - \frac{1}{2\beta}\log\Phi,
\label{CT}
\end{equation} 

and

\noindent where here ${\widetilde{q}}_{ik}$ and $q_{ik}$ are canonical variables, other then the lapse and the shifts functions, in the Einstein and Jordan frames respectively and $\beta$ just a parameter. ${\widetilde{N}}$ $N$, ${\widetilde{N}}_i$ $N_i$ are the lapses and shifts functions respectively in the Einstein and Jordan frames. Therefore the Dirac's constraint analysis, in the Jordan frame, which is a constraint analysis of the Brans-Dicke theory, is performed into the Einstein frame. Here we have just Einstein General Relativity theory minimally coupled to a scalar field. The Dirac's constraint algebra among the secondary first class constraints is just Einstein's geometrodynamics.

Other sources \cite{Deruelle2009} acknowledge the right Hamiltonian transformation from the Jordan to the Einstein frames

\begin{equation}
{\widetilde {h}}_{ab}
 = \phi\,h_{ab}\,,
\quad
{\widetilde {N}}^a
 = N^a\,,
\quad
\widetilde {N}
 = \sqrt\phi\,N\,,
\quad
\widetilde{\phi}
 = \sqrt\frac{3}{2}\,\ln\phi\,.
\end{equation}

\begin{equation}
{\widetilde {p}}^{ab}
 = \frac{1}{\phi}\,p^{ab}\,,
\quad
{\widetilde{\pi}}_{\phi}
 = \sqrt\frac{2}{3}\,(\phi\,\pi_{\phi} - p)\,.
\end{equation}

where ${\tilde{p}}^{ab}$ and $p^{ab}$ are the Hamiltonian momenta conjugated to the three-dimensional metric ${\tilde{h}}_{ab}$ and $h_{ab}$ in the Einstein and Jordan frames respectively. ${\tilde\pi}_{\phi}$ and $\pi_{\phi}$ are the Hamiltonian momenta comjugated to $\tilde{\phi}$ and $\phi$ respectively. $p$ is the trace of $p^{ab}$. When it comes to check the Hamiltonian canonicity of the Hamiltonian transformation from the Jordan to the Einstein frame, they calculate only the Poisson brackets $\{{\widetilde {h}}_{ab},{\widetilde {p}}^{cd}\}_\mathrm J=\{h_{ab},p^{cd}\}_\mathrm J$\,,
$\{{\widetilde {\phi}},\widetilde\pi\}_\mathrm J=\{\phi,\pi\}_\mathrm J$\,,
$\{{\widetilde {p}}^{ab},{\widetilde{\pi}}\}_\mathrm J=0$\,,
$\{\widetilde h_{ab},\widetilde\phi\}_\mathrm J=0$\,,
$\{\widetilde h_{ab},\widetilde\pi\}_\mathrm J=0$\,,
$\{\widetilde p^{ab},\widetilde\phi\}_\mathrm J=0$ (n.b. $\{.,.\}_\mathrm J$ means the Poisson brackets are calculated using the variables of the Jordan frame). They miss the calculations of the Poisson brackets with the lapses and shifts functions, probably because they are considered gauge variables and then without any physical meaning.

In a series of papers, \cite{Gionti2021} \cite{Galaverni:2021jcy} \cite{Galaverni:2021xhd} \cite{GiontiSJ:2023tgx}, we have used the Hamiltonian transformation from the Jordan to the Einstein frames in the case of the Brans-Dicke theory \cite{Brans1961}, in which $f(\phi)=\phi$ and the relative action functional can be found in \cite{Gionti2021},
\begin{eqnarray}
&&\widetilde{N}=\left(16\pi G \phi) \right)^{\frac{1}{2}}N\,, \widetilde{N}_i=\left(16\pi G \phi \right)N_i\,,\nonumber\\
&&\widetilde{h}_{ij}=\left(16\pi G \phi \right)h_{ij}, \nonumber\\
&&{\widetilde \pi}^{ij}=\frac{{\pi}^{ij}}{16\pi G\phi}\,\,\,
{\widetilde \pi}_\phi=\frac{1}{\phi}(\phi \pi_{\phi}-\pi_{h})\,,
\label{tilderelation0}
\end{eqnarray}

\noindent where, as above, the {\it tilde } refers to variables in the Einstein frame. It is very easy to see that 

\begin{equation}
\{{\widetilde N},{\widetilde \pi}_{\phi} \}_J=\frac{8\pi GN}{{\sqrt{16 \pi G\phi}}} \neq 0, \textrm{ and }\; 
\{{\widetilde N}_i,{\widetilde \pi}_{\phi} \}_J=16 \pi G N_i \neq 0\,,
\label{noncanonicalcond}\,\,,
\end{equation}

from which, strictly speaking, the Hamiltonian transformation from the Jordan to the Einstein frames \eqref{tilderelation0} is not a canonical transformation. Therefore, the Dirac's constraint analysis for a scalar-tensor theory (which includes the Brans-Dicke theory) has to be carried out indipendently in the Jordan and the Einstein frames. We have studied the Hamiltonian constrained analysis in Jordan and Einstein frames for the Brans-Dicke theory, a particular case of scalar tensor theory, in both cases $\omega \neq -\frac{3}{2} $ and $\omega =-\frac{3}{2}$, $\omega$ being the coupling constant of the Brans-Dicke theory. In the case $\omega=-\frac{3}{2}$, we have an extra primary first class constraint $C_{\phi}\approx0$, due to the fact that the theory is now also conformal invariant \cite{Galaverni:2021xhd}. The results are summarized by the following two tables 

\begin{table}[ht]
\centering
\begin{tabular}{|c|c|}
 \hline 
  \multicolumn{2}{|c|}{\bf Hamiltonian Analysis of BD for $\omega \neq -\frac{3}{2}$} \\
  {{\bf in Jordan Frame} }    &  {{\bf  in Einstein Frame} }  \\
  \hline
  {{\it constraints}}      & {{\it constraints}}   \\
  {$\pi_N \approx 0; \pi^{i}\approx 0; \mathcal{H}\approx 0;\mathcal{H}_i\approx 0; $}   &  
  {${\widetilde{\pi}_N} \approx 0; {\widetilde{\pi}}_{i}\approx 0; \widetilde{\mathcal{H}}\approx 0;{\widetilde{\mathcal{H}}}_i\approx 0;$}   \\
  \hline
  {{\it constraint algebra }}     &  {{\it constraint algebra}}   \\
    $\{\pi_N,\pi_{i}\}= 0; \{\pi_N,{\mathcal{H}}\}= 0; \{\pi_N,{\mathcal{H}}_i\}= 0; \{\pi_{i},\mathcal{H}\}= 0;$  
  & $\{\widetilde{\pi}_N,\widetilde{\pi_{i}}\}= 0; \{\widetilde{\pi}_N,\widetilde{{\mathcal{H}}}\}= 0; \{\widetilde{\pi}_N,\widetilde{{\mathcal{H}}}_i\}= 0; \{{\widetilde{\pi}}_{i},\widetilde{\mathcal{H}}\}= 0;$ \\
  $\{{\pi}_i,{\mathcal{H}}_j\}= 0;\,\left\{{\mathcal H}(x), {\mathcal H}_i(x')\right\}=-{\mathcal H}(x'){{\partial}'_i}\delta(x,x');$ & 
  $\{{\widetilde{{\pi}}}_i,{\widetilde{{\mathcal{H}}}}_j\}= 0;\,\left\{{\widetilde{{\mathcal H}}}(x), 
  {\widetilde{{\mathcal H}}}_i(x')\right\}=
  -{\widetilde{{\mathcal{H}}}}(x'){{\partial}'_i}\delta(x,x');$ \\
  $\left\{{\mathcal H}_i(x), {\mathcal H}_j(x')\right\} ={\mathcal H}_i(x') \partial_j \delta(x,x')- {\mathcal H}_j(x) {\partial_i}' \delta(x,x');$ & 
  $\{{\widetilde{{\mathcal H}}}_i(x), {\widetilde{{\mathcal H}}}_j(x')\} ={\widetilde{{\mathcal H}}}_i(x') \partial_j \delta(x,x')- {\widetilde{{\mathcal H}}}_i(x) {\partial_i}' \delta(x,x');$\\
  $\{{\mathcal{H}}(x),{\mathcal{H}}(x')\}={\mathcal {H}}^{i}(x)\partial_{i}\delta(x,x')-{\mathcal H}^{i}(x'){\partial}'_{i}\delta(x,x') ;$ & $\{{\widetilde{{\mathcal{H}}}}(x),{\widetilde{{\mathcal{H}}}}(x')\}
  ={\widetilde{{\mathcal H}}}^i(x)\partial_{i}\delta(x,x')-{\widetilde{{\mathcal H}}}^i(x'){\partial}'_{i}\delta(x,x'); $\\
  \hline 
\end{tabular}
\caption{Dirac's constraints and constraint algebra in Jordan and Einstein frames for $\omega\neq -\frac{3}{2}$ (see ref \cite{Gionti2021} for details).}
\label{tab:summary_1}
\end{table}

\begin{table}[ht]
  \centering
\begin{tabular}{|c|c|}
\hline
\multicolumn{2}{|c|}{\bf Hamiltonian Analysis of BD for $\omega = -\frac{3}{2}$}\\
 {{\bf in Jordan Frame} }    &  {{\bf in Einstein Frame} }  \\
  \hline
  {{\it constraints}}      & {{\it constraints}}   \\
  {$\pi_N \approx 0; \pi^{i}\approx 0; C_{\phi}\approx 0; \mathcal{H}^{(-3/2)}\approx 0;\mathcal{H}^{(-3/2)}_i\approx 0; $}   &  
  {${\widetilde{\pi}_N} \approx 0; {\widetilde{\pi}}_{i}\approx 0;{\widetilde{C}}_{\phi}=-\widetilde{\phi}{\widetilde{\pi}}_{\phi}\approx 0; \widetilde{\mathcal{H}}^{(-3/2)}\approx 0;{\widetilde{\mathcal{H}}^{(-3/2)}}_i\approx 0;$}   \\
  \hline
  {{\it constraint algebra }}     &  {{\it constraint algebra}}   \\
  $\{\pi_N,\pi_{i}\}=\{\pi_N,{\mathcal{H}}^{(-3/2)}\}=\{\pi_N,{\mathcal{H}}^{(-3/2)}_i\}=0;$  
  & $\{\widetilde{\pi}_N,\widetilde{\pi_{i}}\}=\{\widetilde{\pi}_N,\widetilde{\mathcal{H}}^{(-3/2)}\}= 0; \{\widetilde{\pi}_N,\widetilde{{\mathcal{H}}}_i^{(-3/2)}\}= 0;$ \\
  $\{\pi_{i},\mathcal{H}^{(-3/2)}\}=\{{\pi}_i,{\mathcal{H}}^{(-3/2)}_j\}=0;$
  &$\{{\widetilde{\pi}}_{i},{\widetilde{\mathcal{H}}}^{(-3/2)}\}=\{{\widetilde{{\pi}}}_i,{\widetilde{{\mathcal{H}}}}^{(-3/2)}_{j}\}=0;$\\
  $\left\{C_{\phi}(x),{\mathcal {H}}_{i}^{(-3/2)}(x')\right\}=-\partial'_{i}\delta(x,x')C_{\phi}(x');$ &$\left\{{\widetilde C}_{\phi}(x),{\widetilde{{\mathcal {H}}}_{i}}^{(-3/2)}(x')\right\}=0;$\\
  $ \left\{C_{\phi}(x),{\mathcal {H}}^{(-3/2)}(x')\right\}=\frac{1}{2}{\mathcal{H}}^{(-3/2)}(x)\delta(x,x');$ & 
   $\left\{\widetilde{C}_{\phi}(x),{\widetilde{{\mathcal {H}}}}^{(-3/2)}(x')\right\}=0;$ \\
  $\left\{{\mathcal H}^{(-3/2)}(x), {\mathcal H}^{(-3/2)}_i(x')\right\}=-{\mathcal H}^{(-3/2)}(x'){{\partial}'_i}\delta(x,x');$ & 
  $\left\{{\widetilde{{\mathcal H}}}^{(-3/2)}(x), 
  {\widetilde{{\mathcal H}}}^{(-3/2)}_i(x')\right\}=
  -{\widetilde{{\mathcal{H}}}}^{(-3/2)}(x'){{\partial}'_i}\delta(x,x');$ \\
  $\left\{{\mathcal H}^{(-3/2)}_i(x), {\mathcal H}^{(-3/2)}_j(x')\right\} ={\mathcal H}^{(-3/2)}_i(x') \partial_j \delta(x,x')$ & 
  $\{{\widetilde{{\mathcal H}}}^{(-3/2)}_i(x), {\widetilde{{\mathcal H}}}^{(-3/2)}_j(x')\} ={\widetilde{{\mathcal H}}}^{(-3/2)}_i(x') \partial_j \delta(x,x')$\\
  $ - {\mathcal H}^{(-3/2)}_j(x) {\partial_i}' \delta(x,x');$ & $- {\widetilde{{\mathcal H}}}^{(-3/2)}_i(x) {\partial_i}' \delta(x,x'); $\\
  $\{{\mathcal{H}}^{(-3/2)}(x),{\mathcal{H}}^{(-3/2)(x')}\}=$ & $\{{\widetilde{{\mathcal{H}}}}^{(-3/2)}(x),{\widetilde{{\mathcal{H}}}}^{(-3/2)}(x')\}
  = $\\
  ${\mathcal {H}}^{(-3/2)}_{i}(x)\partial^{i}\delta(x,x')-{\mathcal H}^{(-3/2)}_{i}(x'){\partial}'^{i}\delta(x,x')+  $ & ${\widetilde{{\mathcal H}}}^{(-3/2)}_i(x)\partial^{i}\delta(x,x')-{\widetilde{{\mathcal H}}}^{(-3/2)}_i(x'){\partial}'_{i}\delta(x,x'); $\\
  $\left[D^{i}(\log\phi(x))\right]C_\phi(x)\partial_{i}\delta(x,x')$ & ${ }$\\
  $-\left[D^{i}(\log\phi(x'))\right]C_\phi(x'){\partial}'_{i}\delta(x,x'); $& ${ }$\\
  \hline 
\end{tabular}
\caption{Dirac's constraints and constraint algebra in Jordan and Einstein frames for $\omega= -\frac{3}{2}$}
\label{tab:summary_2}
\end{table}

The Dirac's constraint algebra is the same in the Jordan and Einstein frame for $\omega\neq -\frac{3}{2}$ (see table \eqref{tab:summary_1}). The case $\omega= -\frac{3}{2}$ looks a bit more complicate \cite{Galaverni:2021xhd} (see table \eqref{tab:summary_2}). The Hamiltonian transformation from the Jordan to the Einstein frames is singular \cite{Galaverni:2021xhd}. Therefore, the Dirac's constraint algebra in the two frames is different (see table\eqref{tab:summary_2}). 

We have studied \cite{Galaverni:2021jcy} a flat Brans-Dicke FLRW mini-superspace model in the Jordan and the corresponding flat Einstein FLRW  minimally coupled to a scalar field in the Einstein frame in the Hamiltonian formalism. We have noticed that if we apply the transformation from the Jordan to the Einstein frames on the Hamiltonian equations of motion, the transformation does not, strictly speaking, map solutions of the equations of motion in one frame into solution of the equations of motion in the other frame (Hamiltonian canonical transformation). Our result suggests that, in order to make the Hamiltonian transformation, from the Jordan to the Einstein frames, Hamiltonian canonical, we need to gauge-fix the lapse function $N$ \cite{Galaverni:2021jcy} and implement this gauge fixing as a secondary constraint. 

We have implemented the previous observation in a following article \cite{GiontiSJ:2023tgx}. The main idea has been to gauge fix the lapse and the shifts functions and implemented the gauge fixing conditions as a secondary Dirac's constraints. Therefore, in the Jordan frame we have

\begin{equation}
N=c(x)\,; N_i =c_i(x)\,; \longmapsto N-c(x)\approx 0\, \; N_i -c_i(x)\approx 0
\label{secondaryconstraintJF}\, ,
\end{equation}

$c(x)$ and $c_i(x)$ are functions of $x$. In the Einstein frame, the previous gauge fixing becomes:

\begin{eqnarray}
&&\widetilde{N}=c(x)\left(16\pi G \phi \right)^{\frac{1}{2}}\,; \widetilde{N}_i =c_i(x)\left(16\pi G \phi \right)^{\frac{1}{2}}; \nonumber\\
&&\longmapsto\widetilde{N}-c(x)\left(16\pi G \phi \right)^{\frac{1}{2}} \approx 0 \, ;\, \, \widetilde{N}_i -c_i(x)\left(16\pi G \phi \right)^{\frac{1}{2}}\label{secondaryconstraintJF}\, .
\end{eqnarray}

This secondary constraints make the primary first class constraints \cite{GiontiSJ:2023tgx} second class. In fact., it is easy to see 

\begin{equation}
\left\{ N(x)-c(x),\pi_N (x')\right\}=\delta(x,x')\,,
\left\{ N_i(x)-c_i(x),\pi^j (x')\right\}=\delta_i^j\delta(x,x')\,,
\label{secondclassJF}
\end{equation}

and 
\begin{equation}
    \left\{ \widetilde{N}-c(x)\left(16\pi G \phi \right)^{\frac{1}{2}},\pi_N (x')\right\}=\delta(x,x')\,,
\left\{\widetilde{N}_i -c_i(x)\left(16\pi G \phi \right)^{\frac{1}{2}},\pi^j (x')\right\}=\delta_i^j\delta(x,x')\,.
\label{secondclassJF}\end{equation}

Following Dirac's algorithm, we can get rid of the second class constraints defining the Dirac's brackets 

\begin{equation}
\{,\}_{DB}=\{\,,\,\}-C^{-1}_{\alpha \beta}\{\,,\varphi_\alpha\}\{\varphi_\beta, \,\}\,\,\, C_{\alpha\beta}\equiv{\varphi_{\alpha},\varphi_\beta}
\label{diracbrackets}
\end{equation}

where $\varphi_\alpha\,\,\, \varphi_\beta$ are second class Dirac constraints. The equations of motion are now calculated using Dirac's brackets. Once we calculate the equations of motion, we can impose strongly the second class constraints and reduce the number of independent canonical variables. On this reduced phase space, the transformation from the Jordan to the Einstein frames is a Hamiltonian canonical transformation \cite{GiontiSJ:2023tgx}. 
\section{Geometrodynamics in Spherical Symmetry in the Jordan and Einstein Frames}

Consider a Lorentian Manifold $(M,g)$, where we consider a $3+1$ decomposition in spherical symmetry $M=\mathbb{R}\times \Sigma_t$, where $\Sigma_t=\mathbb{R}\times S^2$. The ADM metric is 
\begin{eqnarray}
    g_{\mu\nu}dx^\mu dx^\nu&=& - N^2 dt^2 
      +  \Lambda^2 \left(dr+N^r dt \right)^2  + R^2 d\Omega^2 \nonumber \\
    &=& - \left(N^2-\Lambda^2 (N^r)^2 \right) dt^2
      + 2 \Lambda^2 N^r dt dr\nonumber\\ 
      &&+ \Lambda^2 dr^2
      + R^2 d\Omega^2 \,,  
\label{def:ADMss}
\end{eqnarray}

where we assume $-\infty < r < +\infty $. As in Fig.1, the 4-dimension manifold $(M,g_{\mu\nu})$ can be foliated in 3-dimensions space-like sub-manifolds at a given time $(\Sigma_t,h_{ij})$ with normal vector $n^\mu$. The time-like boundaries correspond to a 3-dimension time-like foliation  
    $(B,\gamma_{ab})$ with normal vector $u^\mu$. Note that in general the space-like and the time-like foliations are not orthogonal ($n^\mu u_\mu\neq 0$). $B_t$ is the two dimensional boundary which is intersection of $\Sigma_t$ and $B$. $\sigma_{AB}$ is the two dimensional metric on $B_t$. On $\Sigma_t$ the extrinsic curvature is $K$, while on $B$ is $\Theta$. The extrinsic curvature on $B_t$ is ${}^{(2)}k$ \cite{Galaverni:2025acu}. $\Sigma_t$ has a spherical symmetry, with topology $\Sigma_t = \mathbb{R}\times S^{2}$.
The line element on a fixed sub-manifold $\Sigma_t$ with metric $h_{ij}$ depends only on two functions $\Lambda(r)$ and $R(r)$ \cite{Berger:1972pg,Kuchar:1994zk}:
$h_{ij}dx^i dx^j={\Lambda}^{2}dr^{2}+R^{2}d{\Omega}^{2}\,\,$
where $d\Omega$ is  the line element on the unit sphere, $d\Omega^2\equiv d\theta^2+\sin^2\theta d\varphi^2$    
    \begin{figure}[htb]
    \centering
    \includegraphics[width=0.6\linewidth]{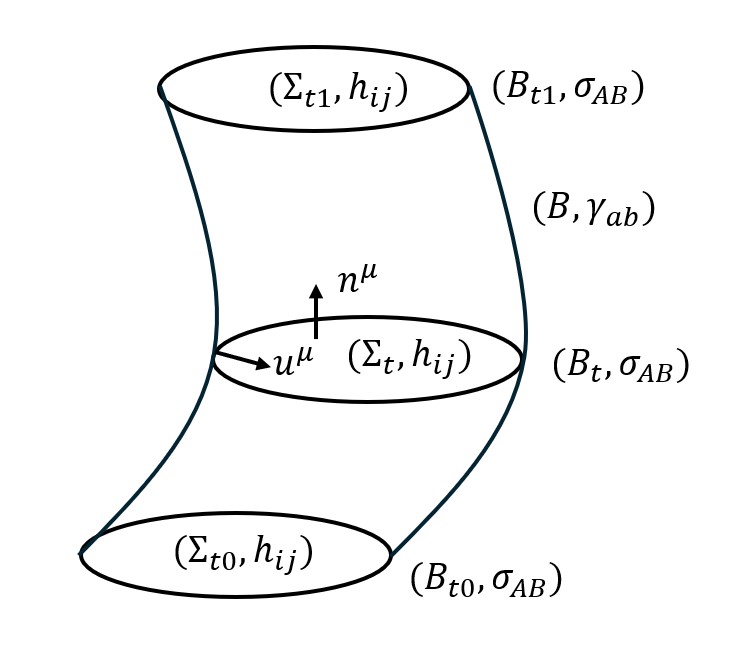}
    \caption{}
    \label{fig:1}
\end{figure}

The Einstein-Hilbert action functional with boundary terms, which make it linear in the variation of the metric tensor, is  \cite{York:1986lje,Lau:1995fr,Hayward:1992ix,Hayward:1993my,Hawking:1996ww,Dyer,Rosabal:2021fao,Reyes:2023sgk}:
\begin{eqnarray}
S &=&\frac{1}{16\pi}
\int_{{M}}d^{4}x \sqrt{-g}\; {}^{(4)}R +\frac{1}{8\pi }\int_{\Sigma_{t_1}}d^{3}x\sqrt{h}K 
-\frac{1}{8\pi }\int_{\Sigma_{t_0}}d^{3}x\sqrt{h}K 
- \frac{1}{8\pi }\int_{B}d^{3}x\sqrt{-\gamma}\,\Theta \nonumber\\
&& - \frac{1}{8\pi }\int_{B_{t_0}} d^2x \sqrt{\sigma} \, \mathrm{arcsinh} (n^\mu u_\mu)  
+\frac{1}{8\pi }\int_{B_{t_1}} d^2x \sqrt{\sigma} \,\mathrm{arcsinh} (n^\mu u_\mu).
 \label{tutta}
\end{eqnarray}

In the Jordan frame, our starting point is the following Lagrangian density

\cite{Callan:1970ze,Bekenstein:1974sf,Bronnikov:2002kf,Faraoni:2004pi,Galtsov:2020jnu,Ray:2024fyx}:  
\begin{eqnarray}
\label{Lagr:JF}
\mathcal{L}_{JF}=\frac{1}{16\pi} \left[\left(1-\frac{\phi^2}{6}\right){}^{(4)}R-g^{\mu\nu}\partial_\mu\phi\partial_\nu\phi \right]\,.
\end{eqnarray}
In this theory the equations of motion for the scalar field $\phi$ are invariant under conformal transformations \cite{Callan:1970ze,Bekenstein:1974sf}.
This model has received attention in the literature because in general relativity and in all other metric theories of gravity, it is the only case for which the equivalence principle holds when $\phi$ is not a gravitational field \cite{Faraoni:2004pi}.
Moreover, this scalar-tensor theory of gravity is, in general, required by the renormalization of the theory \cite{Faraoni:2004pi}. The action, in the Jordan frame, following from \eqref{Lagr:JF}, is similar to \eqref{tutta} extending the considerations in \cite{Dyer} as explained in \cite{Galaverni:2025acu}. In general, this action is defined for compact geometries. For non-compact geometries, we need to define a {\it physical action} $S^{PHYS}$ that is the action above mentioned minus a suitable background geometry $g_0$ \cite{Hawking:1995fd} \cite{Galaverni:2025acu} in such a way the action is not divergent. We can use the definition of the ADM-metric tensor in spherical symmetry \eqref{def:ADMss} to implement the $3+1$ tensor decomposition in the action $S^{PHYS}$ \cite{Galaverni:2025acu}. We can easily calculate the momenta 
\begin{eqnarray}
\pi_\Lambda&\equiv&\frac{\partial \mathcal{L}_{JF}}{\partial \dot{\Lambda}}
=-\frac{1}{N}\left(1-\frac{\phi ^2}{6}\right) R\left(\dot{R}-R^{\prime} N^r\right)+\frac{R^2 \phi}{6 N} (-N^r \phi^\prime+\dot{\phi})\,, \label{eq:PLambda:JF} \\
\pi_R&\equiv&\frac{\partial \mathcal{L}_{JF}}{\partial \dot{R}}=
\frac{1}{N}\left(1-\frac{\phi ^2}{6}\right)\left[R\left(-\dot{\Lambda}+(\Lambda N^r)^\prime\right)+\Lambda\left(-\dot{R}+R^\prime N^r\right)\right]\nonumber\\
&&\qquad\qquad +\frac{\phi \Lambda R}{3 N}\left(\dot{\phi}-N^r \phi^\prime\right)\,,\\
\pi_\phi&\equiv&\frac{\partial \mathcal{L}_{JF}}{\partial \dot{\phi}}=
-\frac{\phi R}{6 N}\left[R\left(-\dot{\Lambda}+(\Lambda N^r)^\prime\right)+2\Lambda\left(-\dot{R}+R^\prime N^r\right)\right]\nonumber\\
&&\qquad\qquad +\frac{\Lambda R^2}{2 N}\left(\dot{\phi}-N^r \phi^\prime\right)\,,
\label{eq:PPhi:JF}
\end{eqnarray}

where $'$ designates the derivative respect to $r$. The ADM canonical Hamiltonian density $\mathcal{H}_{ADM}$ will then be $\mathcal{H}_{ADM}=N \mathcal{H}+N^r \mathcal{H}_r$, where \cite{Galaverni:2025acu} the Hamiltonian constraint $\mathcal{H}$ is 

\begin{eqnarray}
\mathcal{H}&=&\frac{\phi^2\pi_R^2}{36\Lambda}\left(1-\frac{\phi ^2}{6}\right)^{-1}+\frac{\Lambda\pi^2_\Lambda}{2R^2} \left(1+\frac{\phi^2}{18}\right)\left(1-\frac{\phi^2}{6}\right)^{-1}+\frac{\pi_\phi^2}{R^2\Lambda} \left(1-\frac{\phi ^2}{6}\right)
\nonumber\\
&&-\frac{\pi_R\pi_\Lambda}{R}\left(1-\frac{\phi^2}{18}\right)\left(1-\frac{\phi^2}{6}\right)^{-1}+\frac{\phi\pi_R \pi_\phi}{3 R \Lambda}+\frac{\phi \pi_\Lambda\pi_\phi}{3 R^2}\\
&&+\left(1-\frac{\phi ^2}{6}\right)\left(-\frac{\Lambda}{2}+\frac{R^{\prime 2} }{2\Lambda}-\frac{R R^\prime \Lambda^\prime}{\Lambda^2}+\frac{R R^{\prime\prime}}{\Lambda}\right)+\frac{R^2\phi^{\prime 2}}{12 \Lambda}
+\frac{R^2 \phi \Lambda^\prime \phi^\prime}{6\Lambda^2}-\frac{R^2 \phi \phi^{\prime\prime}}{6\Lambda}
-\frac{R\phi R^\prime \phi^\prime}{3\Lambda}\,,\nonumber
\label{HamiltianconJF}
\end{eqnarray}

and the momentum constraint $\mathcal{H}_r$ is

\begin{eqnarray}
\mathcal{H}_r=\pi_{R} R' - \Lambda \pi_{\Lambda}' + \pi_{\phi}{\phi}'\,. 
\label{momentumconstraint:JF}
\end{eqnarray}

From the Hamiltonian it is quite straightforward to derive the equations of motion \cite{Galaverni:2025acu}.

If we apply the following transformation from the Jordan to the Einstein frame, 

\begin{equation}
\widetilde{g}_{\mu\nu}=\left(1-\frac{{\phi}^2}{6}\right)g_{\mu\nu}\,,\,\mathrm{and}\,\,
\widetilde{\phi}=\sqrt{6} \tanh^{-1}\frac{\phi}{\sqrt{6}}\,,
\label{def:conf:mphi}
\end{equation}

we get the following Lagrangian density in the Einstein frame 

\begin{eqnarray}
\mathcal{L}_{EF}=\frac{1}{16\pi} \left({}^{(4)}\widetilde{R}-\widetilde{g}^{\mu\nu}\partial_\mu\widetilde{\phi}\,\partial_\nu\widetilde{\phi} \right)\,,
\label{EF:Lagr}
\end{eqnarray}

from this Lagrangian it is possible to derive, as in the case of the Jordan Frame, the Hamiltonian density and the equations of motion \cite{Galaverni:2025acu}. The relations among the canonical variable in the Jordan and Einstein frames is the following \cite{Galaverni:2025acu}

\begin{eqnarray}
&&\widetilde{N}=\left(1-\frac{{\phi}^2}{6}\right)^{\frac{1}{2}}N\,,\,\,{\widetilde{N}}^r=N^r\,,\,\,\nonumber\\
&&\widetilde{\Lambda}=\left(1-\frac{{\phi}^2}{6}\right)^{\frac{1}{2}}\Lambda\,,\,\,\widetilde{R}=\left(1-\frac{{\phi}^2}{6}\right)^{\frac{1}{2}}R\,.    
\label{def:conf:metric}
\end{eqnarray}

\begin{eqnarray}
\label{EFJF:momenta1}
{\widetilde{\pi}}_{\Lambda}&=&\left(1-\frac{{\phi}^2}{6}\right)^{-\frac{1}{2}}{\pi}_{\Lambda}\,,\,\,\\
{\widetilde{\pi}}_R&=&\left(1-\frac{{\phi}^2}{6}\right)^{-\frac{1}{2}}\pi_R\,,\\
\widetilde{\pi}_{\phi}&=& \left( 1-\frac{{\phi}^2}{6} \right)\pi_{\phi}+\frac{1}{6}R\,\phi\,\pi_{R}+\frac{1}{6}\Lambda\,\phi\,\pi_{\Lambda}\,. \label{EFJF:momenta2}\,   
\end{eqnarray}

It is easy to see that the trasndormation from Jordan to Einstein frame is not Hamiltonian canonical

\begin{eqnarray}
\left\{\widetilde{N},{\widetilde{\pi}}_{\phi} \right\}=-N\left(1-\frac{\phi^{2}}{6}\right)^{\frac{1}{2}} \frac{\phi}{6}\,,\\
\left\{\widetilde{N}_r,{\widetilde{\pi}}_{\phi} \right\}=-N_r \left(1-\frac{\phi^{2}}{6}\right)\frac{\phi}{3}\,.
\end{eqnarray}
As explained above \eqref{diracbrackets}, if we gauge fix $N$ and $N_r$ and implemented these gauge fixing conditions as secondary constraints, we generates second class constraints \cite{Galaverni:2025acu}. Then, we define Dirac's brackets and, finally, impose strongly the second class constraints reducing the dimensions of the phase space. On this reduced phase space, the transformation from Jordan to Einstein frames, is Hamiltonian canonical.  

\section{Static solutions in the Jordan and Einstein Frames}

We consider now two static solutions of the equations of motion in the Jordan and in the Einstein frames. 

A particular static solution of the Einstein General Relativity minimally coupled to a massless scalar field, see the EF Lagrangian defined in 
Eq.~\eqref{EF:Lagr}, was studied by Fisher \cite{Fisher:1948yn} and Janis, Newman and
Winicour \cite{Janis:1968zz,Wyman:1981bd,Virbhadra:1997ie} and is is known as FJNW solution:
\begin{eqnarray}
\label{sol:FJNW:EF1}
d\widetilde{s}^2 &=& -\left(1-\frac{b}{r}\right)^{\gamma} dt^2
                     +\left(1-\frac{b}{r}\right)^{-\gamma} dr^2\nonumber\\
                     &&+r^2 \left(1-\frac{b}{r}\right)^{1-\gamma} d\Omega^2\,,\\
\widetilde{\phi}&=&\sqrt{\frac{1-\gamma^2}{2}}\ln\left(1-\frac{b}{r}\right)\,.
\label{sol:FJNW:EF2}
\end{eqnarray}
The constants $b$ and $\gamma$ are related to the mass of the compact gravitational object ($m$) and
the scalar charge ($q$) \cite{Bekenstein:1975ts} by the following expressions \cite{Nandi:2006ds}:
\begin{equation}
b=2\sqrt{m^2+\frac{q^2}{2}}\,,\,\,\gamma=\frac{2m}{b}\,,    
\end{equation}
it also useful to remember the following relation \cite{Galtsov:2020jnu}: $\gamma=\left(1-\frac{2q^2}{b^2}\right)^{1/2}$,
in general for $q\neq 0$ and real we have $0<\gamma<1$, and the FJNW metric is valid only for $r<b$  \cite{Pal:2022cxb}.
Note that for $q=0$, corresponding to $\gamma=1$, the FJNW solution \eqref{sol:FJNW:EF1} reduces to the Schwarzschild solution.

It easy to verify that the equations of motion in the EF \cite{Galaverni:2025acu} are verified by this FJNW solution static solution if we replace:
\begin{eqnarray}
\label{sol:FJNW:N:EF1}
\widetilde{N}&=&\left(1-\frac{b}{r}\right)^{\gamma/2}\,,\;\; \widetilde{N}^r=0\,,\\
\widetilde{\Lambda}&=& \left(1-\frac{b}{r}\right)^{-\gamma/2}\,,\;\; \widetilde{R}=r\left(1-\frac{b}{r}\right)^{(1-\gamma)/2}\,,\\
\widetilde{\phi}&=& \sqrt{\frac{1-\gamma^2}{2}}\ln \left(1-\frac{b}{r} \right)\,.
\label{sol:FJNW:phi:EF1}
\end{eqnarray}
Note that all momenta vanish since FJNW is a static solution. Notice that on the lapse $\widetilde{N}$ and the shift ${\widetilde{N}}^r$ we have implemented a gauge fixing. 

Applying the transformation from the Jordan to the Einstein frame on the FJNW solution, we can do it since it is a Hamiltonian  canonical trasformation according to \cite{GiontiSJ:2023tgx},   of the EF case, see Eqs.~\eqref{sol:FJNW:N:EF1}-\eqref{sol:FJNW:phi:EF1}, we obtain a solution in the JF. For simplicity, we consider the  $\gamma=1/2$ case. It is, also, useful to introduce a new coordinate $\rho$ defined as:
\begin{equation}
\label{def:rho}
 \sqrt{1-\frac{b}{r}} \equiv   1-\frac{b}{2 \rho}\,.     
\end{equation}
In terms of this new coordinate $\rho$ the canonical variables in the JF are:
\begin{eqnarray}
&&N(\rho)=1-\frac{b}{4\rho}\,,\;\; N^r(\rho)=0\,,\\
&&\Lambda(\rho)=\frac{1}{2}\left(1+\frac{1}{\sqrt{1-\frac{b}{r} }}\right)\frac{dr}{d\rho}= \left(1-\frac{b}{4\rho}\right)^{-1}\,,\\
&& R(\rho)=\rho\,,\;\;\phi(\rho)=-\sqrt{6}\frac{ b/4}{\rho-b/4} \,,
\end{eqnarray}
and the solution takes the BBMB (Bocharova-Bronnikov-Melnikov-Bekenstein) form \cite{Bekenstein:1974sf,Galtsov:2020jnu}: 
\begin{eqnarray}
ds^2&=&-\left(1-\frac{b}{4\rho}\right)^2 dt^2
       +\left(1-\frac{b}{4\rho}\right)^{-2} d\rho^2\nonumber\\
       &&+\rho^2 d\Omega^2\,, \label{BBMB1}\\
\phi&=&-\sqrt{6}\frac{\frac{b}{4}}{\rho-\frac{b}{4}}\,.\label{BBMB2}
\end{eqnarray}
In this case $\rho>0$:
$\rho=b/4$ corresponds to the event horizon of the BBMB solution \cite{Turimov:2025odi}. 
For this particular value of the radial coordinate, the relation between $\rho$ and $r$ is singular, see Eq. \eqref{def:rho} 
\begin{equation}
\frac{r}{b}=\left[1-\left(1-\frac{b}{2\rho}\right)^2\right]^{-1}\,.    
\end{equation}

Notice that, as shown in \cite{Galaverni:2025acu}, the FJNW naked singularity is mapped into a regular point in the Jordan frame as a consequence of the fact that the Weyl(conformal) factor becomes null for this value of the scalar field. 

\section{Conlusion}
We have summarized and discussed several recent results on the Hamiltonian transformation between the Jordan and the Einstein frames. As we have tried to stress, technically speaking, the transformation is not Hamiltonian canonical on the extended phase space where the Dirac's Hamiltonian constraints are not zero. The transformation between the two frames becomes Hamiltonian canonical, provided we gauge fix the lapse and the shifts functions and implement the gauge fixing conditions as secondary Dirac's constraints. We have studied also the spherically symmetric case with the appropriate boundary terms. We have included a known example of the mapping of the FJNW naked singularity solution into BBMB Black Hole, stressing that the transformation from JF to EF can be applied on the solutions of the equations of motion since the transformation itself is Hamiltonian canonical. The physical equivalence between the two frames, as regards the "observables", remains still an open question. Surely, the transformation from the JF to the EF generates new solutions of Einstein General Relativity. 

\section{Acknowledgements}
We thank Matteo Galaverni for collaboration on this research and for carefully reading and correcting this paper.

\bibliographystyle{iopart-num}
\bibliography{bransdickepartcase} 

\end{document}